\documentclass[aps,prl,showpacs,amsmath,amssymb,amsfonts,twocolumn,lengthcheck]{revtex4-1}

\usepackage{graphicx}
\usepackage{subfigure}
\usepackage{verbatim}
\usepackage{dcolumn}
\usepackage{bm}
\usepackage{epsf}
\usepackage{color}
\usepackage[colorlinks=true,citecolor=blue,linkcolor=blue,urlcolor=blue]{hyperref}

\newcommand{\ket}[1]{\left|#1\right\rangle}
\newcommand{\braket}[2]{\left\langle #1|#2\right\rangle}

\newcommand{\tr}[1]{\mathrm{tr}\left\{#1\right\}}

\newcommand{\la}{\left\langle}
\newcommand{\ra}{\right\rangle}
\newcommand{\pd}{\partial}
\newcommand{\de}[1]{\delta\left(#1\right)}
\newcommand{\td}{\mathrm{d}}

\newcommand{\etal}{\textit{et al. }}
\newcommand{\e}[1]{\exp{\left(#1\right)}}
\newcommand{\lo}[1]{\ln{\left(#1\right)}}
\newcommand{\id}{\mathbb{I}}
\newcommand{\com}[2]{\left[#1,\,#2\right]}

\newcommand{\bla}{bla\\bla\\bla\\bla\\bla}

\newcommand{\PRL}{Phys. Rev. Lett. }

\newcommand{\mc}[1]{\mathcal{#1}}

\newcommand{\mrm}[1]{\mathrm{#1}}

\DeclareMathOperator*{\sumint}{%
\mathchoice%
  {\ooalign{$\displaystyle\sum$\cr\hidewidth$\displaystyle\int$\hidewidth\cr}}
  {\ooalign{\raisebox{.14\height}{\scalebox{.7}{$\textstyle\sum$}}\cr\hidewidth$\textstyle\int$\hidewidth\cr}}
  {\ooalign{\raisebox{.2\height}{\scalebox{.6}{$\scriptstyle\sum$}}\cr$\scriptstyle\int$\cr}}
  {\ooalign{\raisebox{.2\height}{\scalebox{.6}{$\scriptstyle\sum$}}\cr$\scriptstyle\int$\cr}}
}

\begin{document}

\title{Jarzynski equality in $\mc{PT}$-symmetric quantum mechanics}

\author{Sebastian Deffner}
\author{Avadh Saxena}
\affiliation{Theoretical Division and Center for Nonlinear Studies, Los Alamos National Laboratory, Los Alamos, NM 87545, USA}

\date{\today}

\begin{abstract}
We show that the quantum Jarzynski equality generalizes to $\mc{PT}$-symmetric quantum mechanics with unbroken $\mc{PT}$-symmetry. In the regime of broken $\mc{PT}$-symmetry the Jarzynski equality does not hold as also the $\mc{CPT}$-norm is not preserved during the dynamics. These findings are illustrated for an experimentally relevant system -- two coupled optical waveguides. It turns out that for these systems the phase transition between the regimes of unbroken and broken $\mc{PT}$-symmetry is thermodynamically inhibited as the irreversible work diverges at the critical point.
\end{abstract}


\maketitle

Over the last two decades $\mc{PT}$-symmetric quantum mechanics has established itself as an important area of modern research. In its original formulation \cite{Bender1998} $\mc{PT}$-symmetric quantum mechanics was a rather mathematical theory with only loose connection to physical reality \cite{Bender2013}. However, recent experimental progress has provided a realization of $\mc{PT}$-symmetric systems, which can be understood as systems with balanced loss and gain \cite{Bender2003,Klaiman2008,Makris2008,Musslimani2008,Bender2013}. In particular, R\"uter \etal observed $\mc{PT}$-symmetry in an optical system \cite{Ruter2010}, which consists of two coupled waveguides with modulated refraction indexes. 

Around the same time of the first description of $\mc{PT}$-symmetric quantum mechanics \cite{Bender1998} Jarzynski achieved a major breakthrough in thermodynamics of small systems \cite{Jarzynski1997}. The Jarzynski equality, $\la\e{-\beta W}\ra=\e{-\beta \Delta F}$, allows to determine the free energy difference for an isothermal process from an ensemble of non-equilibrium realizations of the process. Here, $\beta$ is the inverse temperature, $W$ is the non-equilibrium work, and $\Delta F$ is the free energy difference. The angular brackets, $\la \cdot\ra$, denote an average over all possible work values. Therefore, modern thermodynamics places special interest on the distribution of work values, $\mc{P}(W)$, which has been studied extensively, for instance, in  classical systems \cite{Mazonka1999,Speck2011,Kwon2013a,Hoppenau2013a} as well as for quantum systems \cite{Deffner2008,Huber2008,Talkner2008,Deffner2010a,Quan2012,Talkner2013,Gong2014,Leonard2014,Batalhao2014,An2014b}.

It was shown that the classical Jarzynski equality can be generalized to isolated quantum systems \cite{Kurchan2000,Tasaki2000}, for which the thermodynamic work is determined by a two-time energy measurement \cite{Talkner2007,Campisi2011}. In this approach one commonly considers the following procedure: a quantum system is prepared in a thermal Gibbs-state; then a projective energy measurement is performed, before the system evolves under an externally controlled Hamiltonian, $H_t$; the procedure is concluded by a second, projective energy measurement \footnote{Projective measurement means that a POVM (positive operator valued measure) is applied to the quantum system. The quantum system is ``projected'' into the basis of the observable, see also \cite{Nielsen2010}}.

The present work addresses the following question: Does the quantum Jarzynski equality together with the two-time energy measurements generalize to $\mc{PT}$-symmetric quantum mechanics? We will find that this is, indeed, the case for systems with \emph{unbroken} $\mc{PT}$-symmetry, whereas for \emph{broken} $\mc{PT}$-symmetry the Jarzynski equality does not hold. These findings will then be carefully analyzed and illustrated for the model describing the optics experiment by R\"uter \etal \cite{Ruter2010}. 

We will show that  the phase transition between unbroken and broken regime has a clear signature in the behavior of the irreversible work, $\la W_\mrm{irr}\ra=\la W\ra -\Delta F$. This is in analogous to systems exhibiting quantum phase transitions \cite{Silva2008,Smacchia2013,Fusco2014a}. However, while quantum phase transitions are thermodynamically allowed, the transition from unbroken to broken $\mc{PT}$-symmetry is thermodynamically inhibited.

\paragraph{Fundamentals of $\mc{PT}$-symmetric quantum mechanics.}

We start by briefly reviewing the main properties of $\mc{PT}$-symmetric quantum mechanics. Consider a quantum system with non-Hermitian, but $\mc{PT}$-symmetric Hamiltonian $H$, i.e., $\left[ \mc{PT},H\right]=0$. Here, $\mc{P}$ is the space reflection (parity) operator, and $\mc{T}$ is the time-reflection operator \cite{Bender1998,Bender2002,Bender2003,Bender2007},
\begin{equation}
\label{q01}
\begin{split}
&\mc{P}\,x\,\mc{P}=-x\quad \mathrm{and}\quad \mc{P}\,p\,\mc{P}=-p\\
&\mc{T}\,x\,\mc{T}=x,\quad \mc{T}\,p\,\mc{T}=-p\quad \mathrm{and}\quad \mc{T}\,i\,\mc{T}=-i\,
\end{split}
\end{equation}
where $x$ and $p$ are position and momentum operator, respectively. Since $\mc{T}$ also changes the sign of the imaginary unit $i$, canonical commutation relations such as $\com{x}{p}=i\hbar$ are invariant under $\mc{PT}$. It has been seen \cite{Bender2007,Bender2013} that $\mc{PT}$-symmetric Hamiltonians generally exhibit two parametric regimes: a regime of \emph{unbroken} $\mc{PT}$-symmetry in which all eigenvalues of $H$ are real, and a regime of \emph{broken} $\mc{PT}$-symmetry  for which the eigenvalue spectrum has real and complex parts. 

The major difference between Hermitian and $\mc{PT}$-symmetric quantum mechanics is the definition of the inner product \cite{Bender2002,Bender2007}. For Hermitian Hamiltonians we have,
\begin{equation}
\label{q02}
\braket{\psi_1}{\psi_2}=\psi_1^\dagger \cdot \psi_2\,,
\end{equation}
where, as usual, $\dagger$ denotes conjugate transpose. This inner product, however, yields indefinite norms for the non-Hermitian, but $\mc{PT}$-symmetric case \cite{Bender1998,Bender2002,Bender2003,Bender2007}. This can easily be seen explicitly, for instance, for the system discussed in the second part of our analysis. Therefore, the inner product in $\mc{PT}$-symmetric quantum mechanics is defined in terms of the metric operator $\mc{C}$ as \cite{Bender2002,Bender2007}
\begin{equation}
\label{q03}
\braket{\psi_1}{\psi_2}_\mc{CPT}=\left(\mc{CPT}\psi_1\right) \cdot \psi_2\,.
\end{equation}
In the unbroken regime $\mc{C}$ can be determined from \cite{Bender2002,Bender2007},
\begin{equation}
\label{q04}
\left[\mc{C},H\right]=0 \quad\mathrm{and}\quad \mc{C}^2=\id\,.
\end{equation}
It is worth emphasizing that the time evolution induced by a time-independent Hamiltonian with unbroken  $\mc{PT}$-symmetry is unitary, and hence the norm is preserved, $\braket{\psi_t}{\psi_t}_\mc{CPT}=1$ for all $t$ \cite{Bender2002,Bender2007}. 

The metric tensor $\mc{C}$ reminds us of the charge conjugation operator from field theory \cite{Bender2007}. However, in the present case $\mc{C}$ is not necessarily associated with an observable, but rather defines a physically consistent theory.

\paragraph{Quantum work for unbroken $\mc{PT}$-symmetry.}

In the following, we are interested in processes that are induced by a time-dependent control parameter $\lambda_t$, with $H_t=H(\lambda_t)$. We start by considering a driving protocol for which $H_t=H(\lambda_t)$ is in the regime of unbroken $\mc{PT}$-symmetry at all times. Note that a time-dependent Hamiltonian $H_t$ also induces a time-dependent metric $\mc{C}_t$ as can be seen from its definition \eqref{q04}. It has been recently shown \cite{Gong2013} that then the generalized Schr\"odinger equation reads,
\begin{equation}
\label{q06}
i\hbar\, \pd_t \ket{\psi_t}=\left(H_t+\mc{A}_t\right)\ket{\psi}\,,
\end{equation}
where $\mc{A}_t$ is a time-dependent gauge field, that is necessary to preserve normalization under the time-dependent metric $\mc{C}_t$. It is given by,
\begin{equation}
\label{q07}
\mc{A}_t=-\frac{i\hbar}{2}\, \mc{W}_t^{-1}\,\pd_t\mc{W}_t\,,
\end{equation}
where $\mc{W}_t$ is the transpose of the metric $\mc{C}_t$, i.e., $\mc{W}_t=\mc{C}_t^T$.

Commonly, the thermodynamic work done during  a process of length $\tau$ is determined by a two-time energy measurement \cite{Talkner2007,Campisi2011}: at initial time $t=0$ a projective energy measurement is performed; then the system is allowed to evolve under the generalized time-dependent Schr\"odinger equation \eqref{q06}, before a second projective energy measurement is performed at $t=\tau$.  For a single realization of this protocol the work is given by
\begin{equation}
\label{q08}
W_{\ket{\phi_m}\rightarrow \ket{\phi_n}}=E_{n}(\lambda_{\tau})-E_{m}(\lambda_0)\,,
\end{equation}
where $\ket{\phi_m}$ is the initial eigenstate with eigenenergy $E_{m}(\lambda_0)$ and $\ket{\phi_n} $ with $E_{n}(\lambda_{\tau}) $ denotes the final state. The distribution of work values is given by averaging over an ensemble of realizations of the process, $\mc{P}(W)=\la \de{W-W_{\ket{\phi_m}\rightarrow \ket{\phi_n}}} \ra$, which can be written as \cite{Kafri2012,Leonard2014}
\begin{equation}
\label{q10}
\mathcal{P}(W)=\sumint_{m,n} \de{W-W_{\ket{\phi_m}\rightarrow \ket{\phi_n}}}\,p\left(\ket{\phi_m}\rightarrow\ket{\phi_n}\right).
\end{equation}
In the last equation the symbol $\sumint$ accounts for discrete and continuous parts of the eigenvalue spectrum. Without loss of generality we assume here that the spectrum is discrete, but see also Ref.~\cite{Leonard2014}. 

In Eq.~\eqref{q10} $p\left(\ket{\phi_m}\rightarrow\ket{\phi_n}\right)$ denotes the probability to observe a specific transition $\ket{\phi_m}\rightarrow\ket{\phi_n} $. This probability is given by \cite{Kafri2012,Leonard2014},
\begin{equation}
\label{q11}
p\left(\ket{m}\rightarrow\ket{n}\right)=\tr{\Pi_{n}\, U_{\tau}\, \Pi_{m}\,\rho_0\,\Pi_{m}\, U_{\tau}^\dagger}\,,
\end{equation}
where $\rho_0$ is the initial density operator of the system and $U_{\tau}$ is the unitary time evolution operator, 
\begin{equation}
\label{q12}
U_{\tau}=\mc{T}_> \e{-\frac{i}{\hbar}\,\int_0^{\tau}\td t\,\left(H_t+\mc{A}_t\right)}\,.
\end{equation}
Here, $\mc{T}_>$ is the time-ordering operator, and $\Pi_\nu$ denotes the projector into the space spanned by the $\nu$th eigenstate. For the sake of simplicity we further assume that all spectra are non-degenerate, for which we simply have $\Pi_\nu=\phi_\nu\cdot \left(\mc{C}_t\mc{PT} \,\phi_\nu\right)$ \footnote{Assuming all spectra to be non-degenerate is not a restriction of the generality of our results. For degenerate spectra the projector is simply replaced by $\Pi_\nu=\sum_\mu \phi_\nu (\mu) \cdot \left(\mc{C}_t\mc{PT} \,\phi_\nu (\mu) \right)$, where  $\{\mu\}$ is the set of states corresponding to the energy $E_\nu$.}.  Hence, Eq.~\eqref{q11} can be written as,
\begin{equation}
\label{q13}
\begin{split}
&p\left(\ket{\phi_m}\rightarrow\ket{\phi_n}\right)=\left(\mc{C}_\tau\mc{PT} \phi_n\right)\cdot \left(U_\tau \phi_m\right)\\
&\quad\cdot\left(\mc{C}_0\mc{PT} \phi_m\right)\cdot (\rho_0 \phi_m) \cdot \left(\mc{C}_\tau\mc{PT} \,U_\tau    \phi_m\right)\cdot\phi_n\,.
\end{split}
\end{equation}
Now, let us assume that the system under study was initially prepared in a Gibbs state, namely we have
\begin{equation}
\label{q14}
\rho_0 \phi_m=\left[\e{-\beta E_m}/Z_0\right] \,\phi_m\,,
\end{equation}
where $Z_0=\tr{\e{-\beta H_0}}$ is the partition function. Then, we compute  the average exponentiated work,
\begin{equation}
\label{q15}
\begin{split}
&\la \e{-\beta W}\ra=\int \td W \mc{P}(W)\,\e{-\beta W}\\
&\quad =\sum_{m,n} \e{-\beta E_{n}+\beta E_{m} }\,p\left(\ket{\phi_m}\rightarrow\ket{\phi_n}\right)\,.
\end{split}
\end{equation}
Substituting Eqs.~\eqref{q13} and \eqref{q14} into Eq.~\eqref{q15} and using the $\mc{CPT}$-normalization of the initial eigenstate $\phi_m$, $\left(\mc{C}_0\mc{PT} \phi_m\right) \cdot \phi_m=1$, we have
\begin{equation}
\label{q16}
\begin{split}
&\la \e{-\beta W}\ra=\left(1/ Z_0\right) \sum_{m,n}\e{-\beta E_n}\\
&\quad \times \left(\mc{C}_\tau\mc{PT} \phi_n\right)\cdot \left(U_\tau \phi_m\right) \cdot\left(\mc{C}_\tau\mc{PT} \,U_\tau    \phi_m\right)\cdot\phi_n \,.
\end{split}
\end{equation}
We further employ the $\mc{CPT}$-symmetric partition of the identity, $\id=\sum_\nu \psi_\nu\cdot \left(\mc{C}_t\mc{PT} \,\psi_\nu\right)$, which is invariant under unitary evolution \cite{Bender2007}, and the $\mc{CPT}$-normalization of the final eigenstate, $\phi_n$. Hence, we obtain
\begin{equation}
\label{q17}
\la \e{-\beta W}\ra=Z_\tau/Z_0 =\e{-\beta \Delta F}\,,
\end{equation}
where $F=-1/\beta \lo{Z}$ is the free energy. In conclusion, we have shown that the quantum Jarzynski equality \eqref{q16} remains valid for quantum systems with unbroken $\mc{PT}$-symmetry, and for which the time-evolution  is described by the generalized Schr\"odinger equation \eqref{q06}.

\paragraph{Regime of broken $\mc{PT}$-symmetry.}

What remains is to check, whether Eq.~\eqref{q17} is also valid in the regime of broken $\mc{PT}$-symmetry. Analogous to the unbroken regime  the initial energy eigenstate can always be chosen to be $\mc{CPT}$-normalized. Therefore, the treatment of the broken regime is identical to the above discussion, if we replace the definition of $\mc{C}$ in Eq.~\eqref{q04} by \footnote{C. M. Bender, private communication}
\begin{equation}
\label{q05}
\left\{\mc{C},H\right\}=0\quad\mathrm{and}\quad \mc{C}^2=\id\,.
\end{equation}
Comparing Eq.~\eqref{q04} with Eq.~\eqref{q05} we observe that in the regime of unbroken $\mc{PT}$-symmetry the metric operator $\mc{C}$ commutes with $H$, whereas in the broken regime the metric operator anticommutes with the Hamiltonian $H$ \footnote{Note that in the unbroken regime all eigenvalues are real. In the broken regime the eigenvalues can complex or even imaginary. The anticommutator instead of the commutator corrects for the sign from the imaginary part.}. In addition, the time-evolution in the broken regime ceases to be unitary as the energy spectrum contains a complex part. Therefore, even when including a gauge field in the generalized Schr\"odinger equation \eqref{q06} to account for time-dependent metrics \eqref{q07}, norms are not preserved. In particular, we have $\sum_m \psi_m \cdot\left(\mc{C}_\tau\mc{PT} \,\psi_m\right)\neq\id$, with $\psi_m=U_\tau \phi_m$, if the time-evolution operator, $U_\tau$, is not at least unital \cite{Kafri2012,Albash2013,Rastegin2013a,Rastegin2014}.  A unital map is a trace preserving, completely positive map, under which the identity is preserved, and that can be written as a superposition of unitary maps \cite{Nielsen2010}. 

In conclusion,  it becomes apparent by inspecting Eq.~\eqref{q16} that the quantum Jarzynski equality together with the two-time energy measurement approach does not hold under broken $\mc{PT}$-symmetry.

\paragraph{$\mc{PT}$-symmetric Jarzynski equality in optics.}

\begin{figure}
\includegraphics[width=.48\textwidth]{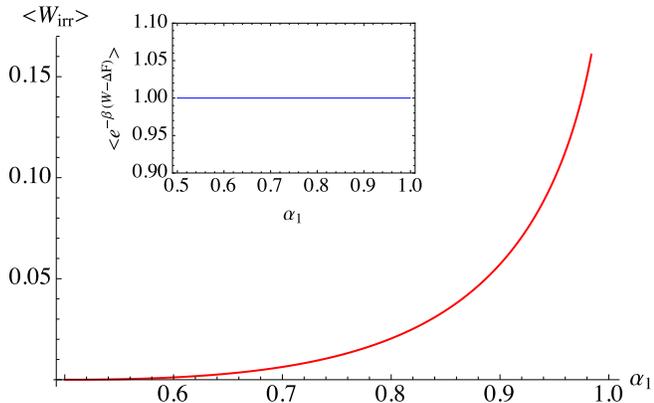}
\caption{\label{fig:unbroken}(color online) Irreversible work $\la W_\mrm{irr}\ra $ \eqref{q28} for the linear protocol \eqref{q27} and $\alpha_0=1/2$, that is a system starting in the regime of unbroken $\mc{PT}$-symmetry. Parameters are $\beta=1$, $\hbar=1$, $\tau=1$, and $\kappa=1$. Inset illustrates the validity of the $\mc{PT}$-symmetric Jarzynski equality \eqref{q17}.}
\end{figure}

The remainder of this discussion is dedicated to a careful analysis and illustration of the above findings for an experimentally relevant example. In a recent experiment R\"uter \etal showed that $\mc{PT}$-symmetric quantum mechanics can be realized in optical set-ups \cite{Ruter2010} -- two coupled waveguides, of which only one is optically pumped. The optical-field dynamics is described by  \cite{Ruter2010} 
\begin{equation}
\label{q18}
\begin{split}
i \,\pd_z E_1&= \frac{i \gamma}{2} \,E_1 - \kappa\, E_2\,,\\
i \,\pd_z E_2&= - \frac{i \gamma}{2} \,E_2 - \kappa\, E_1\,,
\end{split}
\end{equation} 
where $E_{1,2}$ are the field amplitudes in the waveguides, $\kappa$ is the coupling constant, and $\gamma$ is the gain coefficient due to the optical pumping. Identifying the spatial coordinate with a time variable, $z=t/\hbar$, the dynamics of a  $\mc{PT}$-symmetric quantum system is ``frozen'' in the profile of the field amplitudes. In particular, we can identify the Hamiltonian as,
\begin{equation}
\label{q19}
H(\alpha)=\kappa \begin{pmatrix}i\, \alpha & -1 \\-1 &-i \,\alpha \end{pmatrix}\,,
\end{equation}
where we introduced the new parameter $\alpha=\gamma/2\kappa$. Note that this Hamiltonian belongs to the class of two-level systems discussed extensively in the literature \cite{Bender2002,Bender2003,Bender2007}. With the parity operator,
\begin{equation}
\label{q20}
\mc{P}= \begin{pmatrix}0 & 1 \\1 &0 \end{pmatrix}\,,
\end{equation}
and noting that $\mc{T}$ performs here only complex conjugation one easily convinces oneself that $H(\alpha)$ \eqref{q19} is, indeed, $\mc{PT}$-symmetric. The eigenenergies are given by,
\begin{equation}
\label{q21}
\epsilon_{1,2}=\pm \kappa\,\sqrt{1-\alpha^2}\,,
\end{equation}
from which we conclude the regime of unbroken $\mc{PT}$-symmetry, $\alpha\leq 1$, and the broken regime is $\alpha>1$.

It is then a simple exercise to determine the eigenstates and the metric operator $\mc{C}$. We obtain in the unbroken regime for the eigenstates,
\begin{equation}
\label{q22}
\begin{split}
\ket{\phi_1^\mrm{un}}&=\frac{1}{\sqrt{2\sqrt{1-\alpha^2}}}\begin{pmatrix} e^{-\frac{i}{2}\,\mrm{arcsin}( \alpha) }\\ e^{\frac{i}{2}\,\mrm{arcsin} (\alpha) } \end{pmatrix}\,,\\
\ket{\phi_2^\mrm{un}}&=\frac{1}{\sqrt{2\sqrt{1-\alpha^2}}}\begin{pmatrix} e^{\frac{i}{2}\,\mrm{arcsin}( \alpha) }\\ -e^{-\frac{i}{2}\,\mrm{arcsin} (\alpha) } \end{pmatrix}\,,
\end{split}
\end{equation}
with which we obtain
\begin{equation}
\label{q23}
\mc{C}^\mrm{un}= \frac{1}{\sqrt{1-\alpha^2}}\begin{pmatrix}-i\,\alpha & 1 \\1 &i\,\alpha \end{pmatrix}\,.
\end{equation}
Similarly, we have for $\alpha >1$ in the broken regime,
\begin{equation}
\label{q24}
\begin{split}
\ket{\phi_1^\mrm{br}}&=\frac{1}{\sqrt{2\sqrt{\alpha^2-1}}} \begin{pmatrix} e^{\frac{1}{2}\,\mrm{arcosh}( \alpha) }\\ i\,e^{-\frac{1}{2}\,\mrm{arcosh} (\alpha) } \end{pmatrix}\,,\\
\ket{\phi_2^\mrm{br}}&=\frac{1}{\sqrt{2\sqrt{\alpha^2-1}}} \begin{pmatrix} e^{-\frac{1}{2}\,\mrm{arcosh}( \alpha) }\\ i\,e^{\frac{1}{2}\,\mrm{arcosh} (\alpha) } \end{pmatrix}\,,
\end{split}
\end{equation}
and for the metric operator
\begin{equation}
\label{q25}
\mc{C}^\mrm{br}= \frac{1}{\sqrt{\alpha^2-1}}\begin{pmatrix}-i & \alpha \\ \alpha &i \end{pmatrix}\,.
\end{equation}
Finally, we assume the system to be driven externally by varying the parameter $\alpha_t$. This is motivated by the optics experiment of Ref.~\cite{Ruter2010}, where one would change the optical pumping, i.e., vary the gain coefficient, $\gamma$. Such a driving could be implemented in the set-up \cite{Ruter2010} by modulating the refraction index over the length of the waveguide. Projective measurements could be realized by measureing amplitude and phase of the field at both ends of the waveguides.

For broken as well as for unbroken $\mc{PT}$-symmetry the gauge field $\mc{A}_t$ \eqref{q07} reads,
\begin{equation}
\label{q26}
\mc{A}_t=-\frac{i\hbar}{2 \left(\alpha_t^2-1\right)}\,\begin{pmatrix}0&- i\,\dot{\alpha}_t\\  i \,\dot{\alpha}_t &0 \end{pmatrix}\,.
\end{equation}
Equations~\eqref{q19}-\eqref{q26} are all the ingredients necessary to compute the quantum work distribution \eqref{q10} explicitly.

For the sake of simplicity we further assume that $\alpha_t$ is changed linearly from an initial value $\alpha_0$ to a final value $\alpha_1$ during time $\tau$,
\begin{equation}
\label{q27}
\alpha_t=\alpha_0+\left(\alpha_1-\alpha_0\right)t/\tau\,.
\end{equation}
In Fig.~\ref{fig:unbroken} we plot the irreversible work,
\begin{equation}
\label{q28}
\la W_\mrm{irr}\ra =\la W\ra-\Delta F
\end{equation}
as a function of the final value $\alpha_1$ for a system that starts in the regime of unbroken $\mc{PT}$-symmetry -- namely $\alpha_0=1/2$. We observe that $\la W_\mrm{irr}\ra $ is always non-negative, as it should be as an expression of the second law of thermodynamics, and that $\la W_\mrm{irr}\ra $ diverges at the critical point, $\alpha_1=1$. Similar behavior has been observed for quantum phase transitions \cite{Mascarenhas2014}, where the irreversible work can be interpreted as a quantum susceptibility. Diverging susceptibilities are a common feature of quantum critical points \cite{Carr2010}. However, in quantum phase transitions the irreversible work usually exhibits a singularity, i.e., after passing through the transition $\la W_\mrm{irr}\ra $ drops back to a finite value \cite{Mascarenhas2014,Carr2010}. In the present case $\la W_\mrm{irr}\ra $ diverges, signifying a quantum critical point, but stays at infinity even after passing through the phase transition. In other words, the phase transition is thermodynamically ``inhibited'', as the system's response ``freezes out''.  The inset illustrates the validity of the quantum Jarzynski equality for systems with unbroken $\mc{PT}$-symmetry \eqref{q17}

\begin{figure}
\includegraphics[width=.48\textwidth]{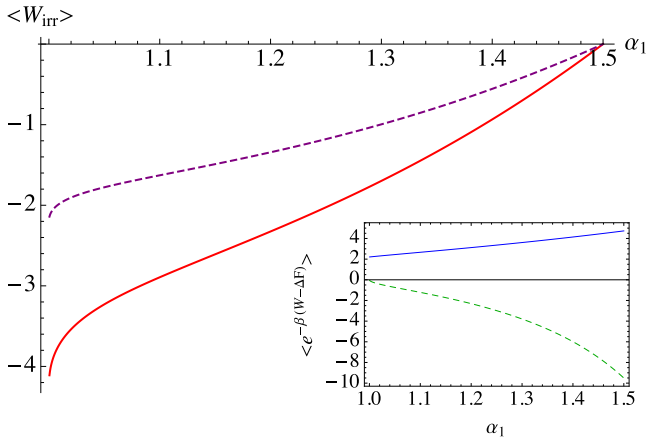}
\caption{\label{fig:broken} (color online) Real part (red, solid line) and imaginary part (purple, dashed line) of the irreversible work $\la W_\mrm{irr}\ra $ \eqref{q28} for the linear protocol \eqref{q27} and $\alpha_0=3/2$, that is a system starting in the regime of broken $\mc{PT}$-symmetry.  Parameters are $\beta=1$, $\hbar=1$, $\tau=1$, and $\kappa=1$. Inset illustrates the violation of the Jarzynski equality \eqref{q17} in real part (blue, solid line) and imaginary part (green, dashed line). }
\end{figure}

Figure~\ref{fig:broken} shows the same quantities, but for a system starting in the regime of broken $\mc{PT}$-symmetry, here $\alpha_0=3/2$. The first observation is that $\la W_\mrm{irr}\ra$ has real and imaginary parts. This is no surprise as the quantum work is defined as an average over differences of eigenenergies \eqref{q08}. In the broken regime, however, the eigenengeries are complex \eqref{q21}, and thus also $\la W_\mrm{irr}\ra $ has to be complex. This signifies the failure of the two-time measurement approach to describe the thermodynamics of systems with broken $\mc{PT}$-symmetry. Nevertheless, we also observe that the absolute value of the irreversible work diverges at the critical point, $\alpha_1=1$, and that also the transition from broken into unbroken regime is thermodynamically inhibited. Finally, the inset illustrates the (complex) violation of the quantum Jarzynski equality, which is in full agreement with the general theory above -- the quantum Jarsynki equality does not hold if the dynamics is not unital \cite{Kafri2012,Albash2013,Rastegin2013a,Rastegin2014}.

\paragraph{Concluding remarks.}

In the present work we have shown how the quantum Jarzynski equality generalizes to $\mc{PT}$-symmetric quantum mechanics. We have found that for quantum systems with unbroken symmetry the Jarzysnki equality holds, while this is not the case in the broken regime. The crucial requirement is that the dynamics is at least unital. Hence, in time-dependent $\mc{PT}$-symmetric quantum mechanics the Schr\"odinger equation has to be generalized including a gauge field so that all norms are preserved during the driving. 

These findings have been further analyzed for an experimentally relevant system. We have illustrated that the $\mc{PT}$-symmetric Jarzynski equality could be experimentally studied in optical set-ups consisting of  two coupled waveguides. In these systems the time-dependent dynamics is frozen in the amplitude profile of the fields. Time-dependent driving could be implemented by modulation of the refraction index over the length of the waveguide. For this system  we have found that the phase transition between the regimes of unbroken and broken $\mc{PT}$-symmetry is thermodynamically inhibited as the irreversible work diverges at the critical point -- the system's response  ``freezes out''. 

We emphasize that we chose the optical system due to its mathematical simplicity. However, all reported findings are completely general and also apply, for instance, to $\mc{PT}$-symmetric systems in microwave billiards \cite{Dietz2011}, photonic lattices \cite{Regensburger2012},  LRC circuits \cite{lin2012},  optical lattices \cite{Regensburger2013},   metamaterials \cite{Feng2013}, or  phonon lasers \cite{Jing2014}.

\begin{acknowledgements}
We thank C. M. Bender for insightful discussions. SD acknowledges financial support by the U.S. Department of Energy through a LANL Director's Funded Fellowship.
\end{acknowledgements}


%

\end{document}